\documentclass[twocolumn,showpacs,preprintnumbers,amsmath,amssymb]{revtex4}
\usepackage{graphicx}
\usepackage{dcolumn}
\usepackage{bm}

\newcommand{\ie}{{\it i.e.}}

\newcommand{\half}{\frac{1}{2} }

\newcommand{\bfa}{{\cal A}\!\!\!\!\!{\cal A}}

\def\ave#1{\langle~#1~\rangle}
\def\deriv#1{\frac{\partial}{\partial#1}}

\begin{document}

\title{SU(2)$\times$U(1) unified theory for charge, orbit and spin currents}

\author{Pei-Qing Jin$^1$, You-Quan Li$^1$, Fu-Chun Zhang$^{1,2}$}
\affiliation{$^1$Zhejiang Institute of Modern Physics
and Department of Physics, Zhejiang University, Hangzhou 310027, P. R. China\\
$^2$ Centre of Theoretical and Computational Physics and
Department of Physics, the University of Hong Kong, Hong Kong}

\begin{abstract}
Spin and charge currents in systems with Rashba or Dresselhaus
spin-orbit couplings are formulated in a unified version of
four-dimensional SU(2)$\times$U(1) gauge theory, with U(1) the
Maxwell field and SU(2) the Yang-Mills field. While the bare spin
current is non-conserved, it is compensated by a contribution from
the SU(2) gauge field, which gives rise to a spin torque in the
spin transport, consistent with the semi-classical theory of
Culcer et al. Orbit current is shown to be non-conserved in the
presence of electromagnetic fields. Similar to the Maxwell field
inducing forces on charge and charge current, we derive forces
acting on spin and spin current induced by the Yang Mills fields
such as the Rashba and Dresselhaus fields and the sheer strain
field. The spin density and spin current may be considered as a
source generating Yang-Mills field in certain condensed matter
systems.
\end{abstract}

\pacs{72.25.-b, 72.10.-d, 03.65.-w}

\received{\today}

\maketitle

\section{Introduction}\label{introduction}

Spintronics or spin based electronics offers opportunities for a new generation of
electronic devices for information process and storage \cite{Wolf, Zutic}.
One of the recent developments in this field is the study of  the spin Hall effect,
which has potential applications to generate and manipulate spin polarization and spin currents
by applying an electric field~\cite{Hirsch,Zhang,Niu0403,Kato,Wunderlich,Rashba,Hu,
Schliemann,Niu0407,Shen2003,Shen,Inoue,Halperin,ZhangYang}.
The intrinsic transverse spin current induced by an electric field was predicted
by Murakami et. al. ~\cite{Zhang}
in the Luttinger Hamiltonian and by Sinova et al. ~\cite{Niu0403}
in two-dimensional electron systems with a Rashba spin-orbit coupling.
A resonant spin Hall conductance was also predicted by Shen et al. \cite{Shen}
in the latter system when a strong perpendicular magnetic field is applied.
It has been generally agreed by now that the spin Hall conductivity
vanishes in Rashba systems with impurities in the absence of a magnetic field
~\cite{Inoue,Halperin}.
On the experimental side, the coherent spin manipulation and
the electrically induced spin accumulation have been observed~\cite{Kato,Awschalom,Wunderlich}.

One of the current issues in study of spin Hall effect is the non-conservation of the bare
spin current in systems with spin-orbit interaction, which is unfamiliar to the community and
requires better  understanding and interpretation.
Recently, Culcer et al. ~\cite{Niu0407} have developed a semi-classical theory of
spin transport and introduced a torque dipole moment to the spin current.
In this paper, we present a unified SU(2)$\times$U(1) theory for spin and
charge currents. In our theory, both the Rashba and Dresselhaus interactions
are described by an SU(2) Yang-Mills field, and the total spin current contains an additional
contribution from the field strength tensor of the Yang-Mills field,
and is conserved. Our theory  provides
a microscopic understanding of the
spin torque introduced in  the semi-classical spin transport theory.
The orbital-angular-momentum current, or the orbit current
is shown to be non-conserved in the presence of an electromagnetic U(1) field.
We also derive
the forces induced by the Yang Mills field such as the Rashba and
Dresselhaus fields and the sheer strain field
acting on spin and spin current. Finally we argue that the spin density and spin current
may be regarded as a source to generate Yang-Mills fields.

\section{Partially conserved spin current}\label{sec:spincurrent}

It is well known that the Dirac equation for an electron moving in an external potential
$V$ (as well as the Maxwell vector potential $\mathbf{A}$) in
the non-relativistic limit up-to the order of $1/c^2$  leads to the Hamiltonian,
\begin{eqnarray}\label{eq:Hamiltonian-SU(2)}
H &=& \frac{1}{2m}\bigl(\mathbf{p}-\frac{e}{c}\mathbf{A}+
 \frac{2m}{\hbar^2}\hat{\mathbf s}\times\vec{\lambda}\bigr)^2
  +\frac{1}{2}\nabla\cdot\vec{\lambda}
    \nonumber\\
  & & -\frac{2}{\hbar}\mu_{\footnotesize B}^{~} \hat{\mathbf s}\cdot\mathbf B
       +\frac{m\lambda^2}{4\hbar^2} + V.
\end{eqnarray}
where $\vec{\lambda}=\frac{\hbar^2}{4m^2c^2} \nabla V$,
 $\mathbf B =\nabla\times \mathbf{A}$,
 and $e$ denotes the charge of the carriers under consideration.

The first term in Eq. (\ref{eq:Hamiltonian-SU(2)}) indicates
the Rashba spin-orbit coupling ({\it e.g.}, $\vec{\lambda}=\alpha \hat{z}$),
which represents the dynamical momentum involving the interaction of an electron
with both the U(1) Maxwell field and SU(2) Yang-Mills field;
the second is the Darwin term \cite{Darwin},
the third term is the Zeeman energy,
and the last one is a higher order term \cite{Davydov}.

For an electron system confined in the $x$-$y$ plane,
the strength vector of the
Rashba coupling is given by $\vec{\lambda}=b\ave{\mathbf{E}}$ \cite{Nitta9702},
with $\mathbf{E}$ the electric field along the $z-$direction
and $b$ the linear coefficient.
The component of the SU(2) Yang-Mills field potentials
in this system can be expressed in terms of the U(1) Maxwell fields
$\mathbf E$ and $\mathbf B$, hence the SU(2) Yang-Mills field strength
is related to the Maxwell-field strength  and their derivatives.
This appears to be a realistic example of Yang-Mills field
in condensed matter system.

To start with general formalism, let us consider the Schr\"odinger
equation for a particle
moving in an external U(1) Maxwell field and an SU(2) Yang Mills gauge field,
\begin{eqnarray}\label{eq:schroedinger}
&& i\hbar\deriv{t}\Psi(\mathbf{r},~t)=H \Psi(\mathbf{r},~t)\nonumber\\
&& H=\frac{1}{2m}\bigl(\hat{\mathbf p}
 -\frac{e}{c}\mathbf{A} - \eta\bfa^a\tau^a\bigr)^2
  + eA_0 + \eta\mathcal{A}_0^a\tau^a
\end{eqnarray}
where $\Psi$ is a two-component wavefunction. Clearly, in comparison to Eq.~(\ref{eq:Hamiltonian-SU(2)}), $A_0$
and $\mathcal{A}^a_0$ refers to $\frac{m\lambda^2}{4e\hbar^2}+\frac{V}{e}$ and
$-\frac{2}{\hbar}\mu_{\footnotesize B}^{~} \hat{\mathbf s}\cdot\mathbf B$, respectively,
 but our further discussion are valid for the general case.
In this paper, the Greek superscripts/subscripts stand for
$0, 1, 2, 3$, the Latin ones for $1, 2, 3$,
and repeated indices are summed over.
Let us denote the vector potential of the Maxwell
electromagnetic field  by
$A_\mu = (A_0, A_i)$,
and the vector potential of the Yang-Mills field~\cite{Yang} by
$\mathbb{A}_{\mu}=\mathcal{A}_\mu^a\tau^a$,
with $\tau^a$ the generators of SU(2) Lie group.
Similar to the field-strength tensor of the electromagnetic field
(Maxwell field), which is defined by
$F_{\mu\nu}=\partial_\mu A_\nu -\partial_\nu A_\mu$, the field strength tensor of
the Yang-Mills field is defined by
$\mathbb{F}_{\mu\nu}=\mathcal{F}_{\mu\nu}^a\tau^a$,
 whose components are given by
$
\mathcal{F}_{\mu\nu}^a
 =\partial_\mu\mathcal{A}_\nu^a-\partial_\nu\mathcal{A}_\mu^a
 -\eta\epsilon^{abc}\mathcal{A}_\nu^b\mathcal{A}_\mu^c.
$

Let $\hat s^a=(\hat s^1, \hat s^2, \hat s^3)$ be the spin
operators, then the local spin density is given by
$\sigma^a(\mathbf{r},\,t)=\Psi^\dagger(\mathbf{r},\,t) \hat{s}^a \Psi(\mathbf{r},\,t)$.
Using the method similar to the derivation of the continuity equation for the charge current,
 one obtains a \emph{continuity-like} equation
\begin{equation}\label{eq:spin-continuity}
\bigl(\deriv{t}-\eta\vec{\mathcal A}_0\times\bigr)\vec{\sigma}(\mathbf{r},~t)
 +\big(\deriv{x_i}+\eta\vec{\mathcal A}_i\times\bigr)
  \vec{J}_i(\mathbf{r},~t) =0
\end{equation}
where the spin-current density is defined as
\begin{equation}\label{eq:spincurrent}
\vec{\mathbf J}(\mathbf{r},~t)
 = \frac{1}{2}\Re{e}\Psi^\dagger(\vec{s} \mathbf{v} + \mathbf{v} \vec{s})\Psi
\end{equation}
with the velocity operator $\hat{\mathbf v}$ given by
\begin{equation}\label{eq:Heisenberg}
\hat{\mathbf v}=\frac{d\mathbf{r}}{dt}
 =\frac{1}{i\hbar}[\,\mathbf{r},\,H\,]=
 \frac{1}{m}\bigl(\hat{\mathbf p}-\frac{e}{c}\mathbf{A}
 -\eta\bfa^a \tau^a\bigr)
\end{equation}
and $\eta\tau^a=\hat{s}^a$ which are just the spin operators if $\eta=\hbar$.
In order to avoid ambiguities, the indices $i,~j,~k,~\mu,~\nu$
refer to the components of a vector in spatial space
while $a,~b,~c,~\alpha$ refer to those in the intrinsic space (Lie algebra space, or spin space).
Moreover, a vector in spatial space is denoted by a bold
face while that in intrinsic space is specified by an overhead arrow, {\it e.g.}, $\vec \sigma =(\sigma^1,
\sigma^2, \sigma^3)$, $\mathbf{A}=(A_1, A_2, A_3)$, $\vec{\mathcal A}_i=({\mathcal A}_i^1, {\mathcal A}_i^2,
{\mathcal A}_i^3 )$, $\vec{\mathbf J}=({\mathbf J}^1, {\mathbf J}^2, {\mathbf J}^3)$ and ${\mathbf J}^3=(J^3_1,
J^3_2, J^3_3)$ etc..

Let us discuss the physics implications of Eq.~(\ref{eq:spin-continuity}).
In the static case $\vec{\mathbf J}=0$, Eq.~(\ref{eq:spin-continuity}) reduces to
$\partial_t\vec{\sigma}=\eta\vec{\mathcal A}_0\times\vec\sigma$,
which implies that the $0$-th component of the Yang-Mills field $\vec{\mathcal A}_0$
induces a torsion on the local spin density $\vec{\sigma}$. From Eq. (3), we can also see that the spin current
is non-conserved even in the limit $\vec{\mathcal A}_0=0$ provided that the spatial components $\vec{\mathcal
A}_i\neq 0$. This is because that the Yang-Mills field $\vec{\mathcal A}_i$ produces a torsion
$\eta\vec{\mathcal A}_i\times\vec{J}_i$ on the spin current, which results in an additional spin precession.

Note that the continuity-like equation for the partially conserved spin current
Eq.~(\ref{eq:spin-continuity}) can be written in a matrix form,
\begin{eqnarray}
\Bigl(\deriv{t}-\Lambda_0\Bigr)
  \begin{pmatrix}
  \sigma^1\\[1mm]
  \sigma^2 \\[1mm]
  \sigma^3
  \end{pmatrix}
 +\Bigl(\deriv{x_i}+\Lambda_i\Bigr)
  \begin{pmatrix}
  J_i^1\\[1mm]
  J_i^2 \\[1mm]
  J_i^3
  \end{pmatrix}=0
\end{eqnarray}
with $\Lambda_\mu=\eta\mathcal{A}_\mu^a\ell^a$. Here $\ell^a$ stand for a representation matrices of the
generators of SO(3) Lie group, namely,
\begin{equation*}{\small
\ell_1\!=\!\begin{pmatrix}
           0 & 0 & 0 \\
           0 & 0 & 1 \\
           0 & -1 & 0
           \end{pmatrix},~
\ell_2\!=\!\begin{pmatrix}
           0 & 0 & -1 \\
           0 & 0 & 0  \\
           1 & 0 & 0
           \end{pmatrix},~
\ell_3\!=\!\begin{pmatrix}
           0  & 1 & 0 \\
           -1 & 0 & 0 \\
           0  & 0 & 0
           \end{pmatrix}. }
\end{equation*}
This provides a geometric picture that the non-conservation of spin current implies
the existence of a non-trivial connection characterizing parallel displacements in
space-time manifold, {\it i.e., } the local intrinsic frame rotates from point to point.

\section{Unified operators for charge and spin currents}\label{sec:operator}

As we shall see below, it will be instructive to consider spin and charge in a four-dimensional intrinsic space
with three dimensions in spin and one in charge. The continuity equation related to the conventional charge
density $\rho = e\Psi^\dag\Psi$ reads
\begin{equation}\label{eq:continuity}
\frac{\partial\rho}{\partial t} + \nabla\cdot\mathbf{j}=0,
\end{equation}
where ${\mathbf j}= e\Re{\mathrm e}\Psi^\dag\hat{\mathbf v}\Psi$ is the charge current density. We introduce a
four-dimensional velocity and a four-dimensional current valued on the U(1)$\times$SU(2) gauge group. Let
$\tau^\alpha=(\tau ^0,\,\tau^a\,)$ with $\tau^0$ being the identity matrix, the four-dimensional velocity
operator is defined as $\hat{v}_{\mu}= v_ \mu ^\alpha \tau^\alpha$, which is determined by the Heisenberg
equation of motion,
\begin{equation}
\hat v_ \mu =\frac{1}{i\hbar}[x_{\mu},H].
\end{equation}
In the above equation, we have taken the commutation relation between $x_0=ct$
 and the Hamiltonian to be $ic\hbar$.
For the Hamiltonian (\ref{eq:schroedinger}), we can write
\begin{equation}\label{eq:4D-velocity}
\hat v _\mu \equiv (\hat v_0,\,\hat v _i )=(c,\,v_i^0+ v_i^a \tau^a ),
\end{equation}
with $v_0^0=1$, $v_0^a=0$; $v_i^0=(\hat p_i -\frac{e}{c}A_i)/m$ and $v_i^a =-\eta\mathcal{A}_i^a/m$. Using these
velocity operators, a unified four-dimensional "current tensor operator" including both charge and spin degrees
of freedom, is then defined as:
\begin{equation}\label{eq:currenttensor}
\hat{J}_\mu ^\alpha \equiv \half \{\hat v_\mu,~ \hat s^\alpha \}.
\end{equation}
The anti-commutator guarantees the defined tensor to be hermitian. To make the stipulation consistent with what
we have discussed above, $\alpha=a$ with $a=1,2,3$ refer to spin and $\alpha=0$ refers to charge. Accordingly,
$\hat{s}^0 = e\tau^0$. From Eq.(\ref{eq:currenttensor}) we obtain, for $\mu=0$,
\begin{equation}
\hat{J}^\alpha_0=
\left\{\begin{array}{l}
          \hat{J}_0^0 = ec\tau^0,\\[1mm]
          \hat{J}_0^a = c\hat{s}^a
       \end{array}
\right.
\end{equation}
which recovers the usual definitions of charge and spin densities,
\begin{eqnarray}
\rho(\mathbf{r})&\equiv& \frac{1}{c}J^0_0(\mathbf{r})
 =\frac{1}{c}\Psi^\dagger\hat{J}^0_0\Psi = e\Psi^\dagger\Psi,
  \nonumber\\
\sigma^a(\mathbf{r})&\equiv& \frac{1}{c}J^a_0(\mathbf{r})
 =\frac{1}{c}\Psi^\dagger\hat{J}^a_0\Psi=\Psi^\dagger\hat{s}^a\Psi.
\end{eqnarray}
For $\mu=i$, we have
\begin{eqnarray*}
\hat{J}^\alpha_i &=& \half \{\hat{v}_i,\hat{s}^\alpha\}
 =\left\{
  \begin{array}{l}
               \hat{J}_i^0 = e\hat{v}_i  \\[1mm]
               \hat{J}_i^a = \half(\hat{v}_i\hat{s}^a +\hat{s}^a\hat{v}_i)
               \end{array}
        \right.
\end{eqnarray*}
which defines the charge and spin current densities,
\begin{eqnarray}
j_i(\mathbf{r})&\equiv & J^0_i(\mathbf{r})
  =\Re{\mathrm e}\Psi^\dagger\hat{J}^0_i\Psi=
   e\Re{\mathrm e} \Psi^\dagger \hat{v}_i\Psi,
    \nonumber\\
J^a_i(\mathbf{r})&\equiv & J^a_i(\mathbf{r})
 =\Re{\mathrm e}\Psi^\dagger\hat{J}^a_i\Psi
  =\Re{\mathrm e} \Psi^\dagger\half(\hat{v}_i\hat{s}^a + \hat{s}^a \hat{v}_i)\Psi.
   \nonumber\\
\end{eqnarray}
The continuity-like equations (\ref{eq:spin-continuity})
can be written in a four-dimensional form
\begin{equation}\label{eq:4D-continuity}
\deriv{x_\mu}\mathbb{J}_\mu
  +i\eta\bigl[\mathbb{A}^\mu,\,\mathbb{J}_\mu \bigr] = 0,
\end{equation}
where $\mathbb{J}_\mu=J^a_\mu \tau^a$ denotes the spin-current vector
which is a matrix valued vector, and
$\mathbb{A}^\mu$ is obtained from $\mathbb{A}_\mu$ by
the Minkowski metric tensor $g^{\mu\nu}=\mathrm{diag}(1~,-1,~-1,~-1)$.
This shows that Eq. (\ref{eq:spin-continuity}) is an SU(2) covariant extension
of traditional continuity equation.
By comparing the above equation with the continuity equation for the
charge current, we see that
the non-vanishing terms in Eq. (\ref{eq:spin-continuity}), or
Eq.~(\ref{eq:4D-continuity}) in systems with Rashba spin-obit coupling is just the
spin torque in the semi-classical theory
of Culcer et al.~\cite{Niu0407}.  The spin torque
represents the spin procession caused by an external magnetic field
and the Rashba interaction.
Since the rotational symmetry in spin space is broken,
the spin density is not conserved.
In  our theory the non-conservation of spin current in the presence
of Yang-Mills field is due to its non-Abelian feature.

Since the spin current defined in Eq.~(\ref{eq:spincurrent}) is
non-conserved, it will be interesting to examine the origin of the
non-conservation by means of constructing a conserved "total
current".

\section{Conservation of a total spin current}

An electron has two important intrinsic properties:
its charge and spin. As we have illustrated, the Rashba or Dresselhaus coupling describes
the interaction between an electron and some particular SU(2) gauge fields.
The SU(2) gauge potential was also adopted in the discussion of the
quantum interference of a magnetic moment in magnetic fields \cite{Anandan}.
We are therefore motivated to construct a theory with the gauge field
coupled to spin and spin current in Lagrangian formalism.

The non-relativistic Lagrangian density of the system in the symmetric form is given by
\begin{eqnarray}\label{eq:Lagrangian}
&&\mathcal L_{NR}=
\frac{i}{2}(\dot\Psi^\dag\Psi-\Psi^\dag\dot\Psi)+\Psi^\dag(eA_0 + \eta\mathcal A_0^a)\Psi \nonumber \\
&&+\frac{1}{2m}\bigl[(\mathbf p -\frac{e}{c}\mathbf A-\eta\bfa^a\tau^a)\Psi\bigr]^\dag
\cdot\bigl[\mathbf p - \frac{e}{c} \mathbf A-\eta\bfa^a\tau^a\bigr]\Psi   \nonumber \\
&&-\frac{1}{4}F_{\mu\nu}F_{\mu\nu} -\frac{1}{4}\mathcal{F}_{\mu\nu}^a \mathcal{F}_{\mu\nu}^a ,
\end{eqnarray}
where $\dot\Psi=\deriv t\Psi$.
According to the Euler-Lagrangian equation, this Lagrangian density gives the
same equation of motion as Eq.~(\ref{eq:schroedinger})
\begin{equation}
i\deriv{t}\Psi=\Bigl[\frac{1}{2m}(\mathbf p - \frac{e}{c}\mathbf A - \eta\bfa^a\tau^a)^2
+ eA_0 + \eta\mathcal A_0^a\tau^a \Bigr]\Psi.
\end{equation}
For the system with a Rashba spin-orbit coupling in some semiconductors, the SU(2) gauge potential is
antisymmetric $\mathcal{A}^a_i=-\mathcal{A}^i_a$, namely,
\begin{equation}\label{eq:lamda}
\mathcal{A}_i^a =\frac{\hbar}{8\eta m c^2}\epsilon^{iab}\partial_b V.
\end{equation}
However, our formalism here in terms of the Yang-Mills field covers more general
cases, such as systems with a Dresselhaus coupling.

Noether's theorem suggests that the invariance of the system under a
continuous transformations will lead to a corresponding
conservation quantity, hence a conserved total current can be defined by
\begin{equation}\label{eq:totalcurrent}
\mathcal{J}_\mu^\alpha = \frac{\delta \mathcal L}{\delta \mathcal{A}^{\mu\alpha} }
  \equiv J_\mu^\alpha + {J^\prime}_\mu^\alpha,
\end{equation}
which obeys
\begin{equation}\label{eq:conserved}
\deriv{x_\mu}\mathcal{J}_\mu^\alpha =0,
\end{equation}
where ${J^\prime}_\mu^\alpha$ refers to the contribution from the Yang-Mills field and Maxwell field. We adopted
a notation $\mathcal{A}^\alpha_\mu =(\mathcal{A}^0_\mu,~\mathcal{A}^a_\mu)= (A_\mu,~\mathcal{A}^a_\mu)$ in
Eq.~(\ref{eq:totalcurrent}) and the explicit components of $J_\mu^\alpha$ can be easily read out as
\begin{eqnarray}\label{eq:jzero}
J^0_0 &=& e\Psi^\dag\Psi, \nonumber \\
J^0_i &=& e\Re{\mathrm e}\Psi^\dag\hat v_i\Psi,
\end{eqnarray}
for the charge case and
\begin{eqnarray}\label{eq:ja}
J^a_0 &=& \eta\Psi^\dag \tau^a\Psi, \nonumber \\
J^a_i &=& \eta\Re{\mathrm e}\Psi^\dag\half(\tau^a\hat v_i+\hat v_i\tau^a)\Psi,
\end{eqnarray}
for the spin case. These results are consistent with the unified four-dimensional
currents we have introduced in previous section.
Eq.~(\ref{eq:jzero}) and Eq.~(\ref{eq:ja}) are just non-relativistic limit of
$\bar{\psi}\gamma_\mu\psi$ and $\bar{\psi}\gamma_5\gamma_\mu\gamma^a\psi$, respectively,
in which
$\psi=\begin{pmatrix} \Psi \\ \chi \end{pmatrix}$ refers to Dirac fermions.

Now let us turn to the extra part ${J^\prime}_\mu^\alpha$. For $\alpha=0$, ${J^\prime}_\mu^0$
is null and the
charge current ${\mathbf j}=e\Re{\mathrm e}\Psi^\dag\hat{\mathbf v}\Psi$ (\ie, $J^0_i$) is just the total
current and hence is conserved, which is consistent with Eq. (\ref{eq:continuity}). Note that the field-strength
tensor of the SU(2) Yang-Mills field $\mathcal{F}_{\mu\nu}^a$ depends on both the derivatives of the potential
$\mathcal{A}_\mu^a$ and the potential itself. This property is very  different from the Maxwell field, which
only contains the derivatives of the potential $A_{\mu}$. Thus the contribution from the Yang-Mills field
${J^\prime}_\mu^a= \eta\epsilon^{abc}\mathcal{A}^b_\nu \mathcal{F}_{\mu\nu}^{c}$ is not null, namely,
\begin{eqnarray}
\vec{J^\prime}_0 &=& -\eta\vec{\mathcal A}_j\times\vec{\mathcal E}_j, \nonumber\\
\vec{J^\prime}_i &=& \eta\vec{\mathcal A}_0\times\vec{\mathcal E}_i +\eta\epsilon_{ijk}\vec{\mathcal
A}_j\times\vec{\mathcal B}_k,
\end{eqnarray}
where ${\mathcal E}$ and ${\mathcal B}$ are the Yang-Mills "electric" and "magnetic" fields. The above density
of (intrinsic) angular momentum ${J^\prime}_0^a$ and its flow ${J^\prime}_i^a$ cancel the non-conservation term
brought about by the bare spin current after substituting them into the continuity equation (\ref{eq:conserved})
for the total current.

The Lagrangian in Eq.~(\ref{eq:Lagrangian}) is shown suitable for
an electron subjected to the Rashba spin-orbit interaction
(\ref{eq:lamda}), while that formalism can be constructed for
other systems with spin-orbit interaction. It is our conjecture
that the non-conservation of the spin current of the electron
system may always be associated with corresponding Yang-Mills
field. In this picture, the Yang-Mills field induced by spin and
spin current carries a fraction of the intrinsic angular momentum,
in analogy to the the Maxwell field induced by charge and charge
current which carries orbital angular momentum.
Finally, we argue
that spin and spin current may be regarded as a source generating
Yang-Mills fields, which in turn interact with the electron. With
the help of Lagrangian formalism similar to
Eq.~(\ref{eq:Lagrangian}), one should be able to study various
physical properties of the system, including the effect of the
Yang-Mills field to the observable. Since the effect of the
Yang-Mills field is weaker than that of the Maxwell field, the
bare spin current has been suggested to be employed in detection
\cite{Stevens}.

\section{Forces acting on spin and spin current}

Using the covariant formalism, we can easily determine the force induced by
the Yang-Mills field on the spin density and spin current (named as spin force for simplicity).
Analogous to the Lorentz force evaluated by
$j_\mu F_{\mu i}$ in electromagnetism, the general form of
the force provided by a Yang-Mills field is given by
\begin{eqnarray}
f_i = J_\mu^a\mathcal{F}_{\mu i}^a = \sigma^a\mathcal{E}_i^a
 - \epsilon_{ijk}J_j^a\mathcal{B}_k^a
\end{eqnarray}
where $\sigma^a$ and $J_i^a$ stand for the spin density and spin current, respectively.

The Yang Mills fields corresponding to the Rashba and Dresselhaus
spin-orbit couplings, the Zeeman term, and the sheer strain field
for a system in $x$-$y$ plane are given by :
\begin{eqnarray}\label{eq:RashbaDresselhaus}
\vec{\mathcal A}_0&=& -\frac{2\mu_{_B}}{\eta}(~B_x, B_y, B_z~),
 \nonumber\\
\vec{\mathcal A}_1&=& \frac{2m}{\eta\hbar}(~\beta,~\alpha,~ \gamma y),
 \nonumber\\
\vec{\mathcal A}_2&=& \frac{2m}{\eta\hbar}(-\alpha, -\beta,~ -\gamma x),
 \nonumber\\
\vec{\mathcal A}_3&=& (0,~0,~0),
\end{eqnarray}
where $\mathbf{B}\equiv (B_x, B_y, B_z)$ stands for the  magnetic
(Maxwell) field, $\alpha$ and $\beta$ refer to the Rashba and
Dresselhaus coupling strengths, and $\gamma$ is related to sheer
strain field, respectively. The "electric" Yang-Mills field is
given by $\mathbb{E}_i=\mathcal{E}_i^a \tau^a$ with
\begin{eqnarray}\label{eq:E-field}
\vec{\mathcal{E}}_i=\frac{2\mu_{_B}}{\eta }\partial_i\vec{B}
  +2\mu_{_B}\vec{\mathcal A}_i \times \vec{B }.
\end{eqnarray}
The "magnetic" Yang-Mills field reads
\begin{eqnarray}\label{eq:B-field}
\mathbb{B}_3=-\frac{2m}{\eta\hbar}(\partial_1\alpha + \partial_2\beta)\tau^1
              -\frac{2m}{\eta\hbar}(\partial_1\beta + \partial_2\alpha)
              \tau^2
                 \nonumber\\
 +\bigl(\frac{4m^2}{\eta\hbar^2}(\beta^2-\alpha^2)
 -\frac{4m}{\eta\hbar}\gamma \bigr)\tau^3, \quad
\end{eqnarray}
while $\mathbb{B}_1=\mathbb{B}_2=0$.
In the above equations, $\alpha$ and $\beta$ are non-uniform in general.
The force can be derived explicitly
\begin{widetext}
\begin{eqnarray}
&& f_1 = \frac{2\mu_B^{}}{\eta}\vec{\sigma}\cdot \partial_1 \vec{B}
   + \frac{4m\mu_B^{}}{\eta\hbar}\bigl[\beta (\vec{\sigma}\times\vec{B})_1
    - \alpha(\vec{\sigma}\times\vec{B})_2 \bigr]
     -J_2^1 \mathcal{B}_3^1 - J_2^2\mathcal{B}_3^2 - J_2^3\mathcal{B}_3^3,
        \nonumber\\
&& f_2 = \frac{2\mu_B^{}}{\eta}\vec{\sigma}\cdot \partial_2 \vec{B}
   - \frac{4m\mu_B^{}}{\eta\hbar}\bigl[\alpha (\vec{\sigma}\times\vec{B })_1
    - \beta(\vec{\sigma}\times\vec{B })_2 \bigr]
     +J_1^1\mathcal{B}_3^1 + J_1^2\mathcal{B}_3^2 + J_1^3\mathcal{B}_3^3,
        \nonumber\\
&& f_3 = \frac{2\mu_B^{}}{\eta}\vec{\sigma}\cdot \partial_3 \vec{B},
\end{eqnarray}
\end{widetext}
The first term in these equations corresponds to the force due to inhomogeneity
of the magnetic field, which is the same as in the Stern-Gerlach apparatus.
Clearly, there are more forces acting on the spin, related to both the
magnetic field  and the
Rashba and Dresselhaus couplings.
Furthermore, the spin current will be subjected to transverse forces.
Note that the inhomogeneous  magnetic field generates a force of Stern-Gerlach type on spin while
the non-uniform Rashba and Dresselhaus coupling introduces a force on the spin current.
In what follows we consider some special cases.

\subsection{Uniform Rashba and Dresselhaus fields}

If the Rashba and Dresselhaus coupling strengths are uniform
($\alpha$ and $\beta$ are constants) and the magnetic field is absent,
we simply have
\begin{eqnarray}\label{eq:RD-force}
f_1 &=& -\frac{2m}{\hbar}(\beta^2 - \alpha^2) J^3_2, \nonumber\\
f_2 &=& \frac{2m}{\hbar}(\beta^2 - \alpha^2) J^3_1, \nonumber\\
f_3 &=& 0.
\end{eqnarray}
Clearly, the forces arising from the Rashba and Dresselhaus couplings
are along the opposite direction.
The magnitudes of the forces are  related to the perpendicular
component ($a=3$) of the spin current only.
Eq.~(\ref{eq:RD-force}) was also obtained in a semi-classical approach~\cite{Shen05}.

\subsection{Non-uniform Rashba and Dresselhaus fields}

If  $\alpha$~\cite{Ohe} and $\beta$ are non-uniform,
there will be a transverse force whose magnitude is
related to the in-plane components ($a=1, 2$) of the spin current.
For example, in the case $|\alpha|=|\beta|$, and $B=0$, we have
\begin{eqnarray}
&& f_1 = \frac{2m}{\eta\hbar}\bigl[
        J_2^1 (\partial_1\alpha + \partial_2\beta)
      + J_2^2 (\partial_1\beta + \partial_2\alpha)\bigr],
        \nonumber\\
&& f_2 = -\frac{2m}{\eta\hbar}\bigl[
      J_1^1 (\partial_1\alpha + \partial_2\beta)
      + J_1^2 (\partial_1\beta + \partial_2\alpha)
       \bigr],
        \nonumber\\
&& f_3 = 0.
\end{eqnarray}

\subsection{Pure sheer strain field}

For electrons subjected to a sheer strain field \cite{Bernevig} only,
which corresponds to $\alpha=\beta=B=0$ but $\gamma\neq 0$,
the forces are given by
\begin{equation}
f_1=\frac{4m}{\eta\hbar}\gamma J_2^3,\quad
f_2=-\frac{4m}{\eta\hbar}\gamma J_1^3,\quad
f_3=0,
\end{equation}
where $\gamma \propto C_3/(\hbar e)$ in the notation of Ref~\cite{Bernevig}.
Clearly, there is a transverse force acting on the spin current.

\section{Orbit density and orbit current}

It is worthwhile to investigate the continuity-like
equation for orbit density and orbit current.
We can define a local density of orbital angular momentum
$\omega^a(\mathbf{r}, t)
 =\half\Psi^\dagger\hat{L}^a\Psi + \half(\hat{L}^a\Psi)^\dagger\Psi.
$
Here $L$ is the dynamical angular momentum,
$\hat{L}^a=\epsilon^{abc}x_b (\hat{p}_c - (e/c)A_c - \eta\mathcal{A}^a\tau^a)$
for the system of an electron moving in a Maxwell field and a Yang-Mills field.
The orbit density so defined is gauge covariant.
Using the Shchr\"odinger equation (\ref{eq:schroedinger}) we can derive
a continuity-like equation
\begin{eqnarray}\label{eq:orbit-continuity}
\deriv{t}\omega^a + \deriv{x_i}I_i^a
   +\epsilon^{abc}x^{}_b\mathcal{F}^\alpha_{c \nu}J^\alpha_\nu=0
\end{eqnarray}
where the flow of the orbital angular momentum, namely the orbit current is given by
\begin{eqnarray}\label{eq:orbit-current}
I^a_i(\mathbf{r}, t)=\Re{\mathrm e}
 \Psi^\dagger\hat{I}^a_i\Psi
   - \frac{1}{4e}\hat{I}^a_i|_{_{A=0}}~\rho
   \nonumber\\
\end{eqnarray}
and
\begin{eqnarray}
\hat{I^a_i}&=&(\hat{v}_i\hat{L}^a + \hat{L}^a \hat{v}_i )/2 \nonumber\\
\hat{v}_i &=&(\hat{p}_i - (e/c)A_i - \eta\mathcal{A}^a\tau^a)/m \nonumber \\
\end{eqnarray}
In Eq.~(\ref{eq:orbit-continuity}) $\mathcal{F}^0_{c i}=F_{c i}$ refers
to the field-strength tensor of the Maxwell field and $\mathcal{F}^a_{c i}$ to
that of the Yang-Mills field; $J^0_i=j^{}_i$ refers to the charge current and
$J^a_i$ the spin current.
Clearly the orbit current (\ref{eq:orbit-current}) so defined
is also gauge covariant.
Note that the definition of the orbit current is not a simple extension
of the spin current (\ref{eq:spincurrent}) by replacing the spin operator
by the orbital angular momentum operator. Actually,
there is one more term in the orbit current, which involves change density $\rho$.

It is worthwhile to point out that the third term in
Eq.~(\ref{eq:orbit-continuity}) is related to
both charge current and spin current.
In the absence of Maxwell and Yang-Mills fields, the third term in
Eq.~(\ref{eq:orbit-continuity}) vanishes and the orbit current is
thus conserved.
Using notations $\vec{\omega}=(\omega_1,~\omega_2,~\omega_3)$,
$\vec{I}_i=(I^1_i,~I^2_i,~I^3_i)$ and
$\vec{r}=(x_1,~x_2,~x_3)$, we can write Eq.~(\ref{eq:orbit-continuity})
in the following form
\begin{eqnarray}\label{eq:the-orbit-continuity}
\deriv{t}\vec{\omega} + \deriv{x_i}\vec{I}_i = \vec{r}\times \vec{F}+\vec{r}\times \vec{f}
\end{eqnarray}
The physics implication of the right-hand side represents torque
produced by Lorentz force $\vec{F}$ due to Maxwell field and the
aforementioned spin force $\vec{f}$ due to Yang-Mills field.

\section{Summary}

We have introduced a four-dimensional charge and spin current tensor
for systems coupled with Yang-Mills fields.
The Rashba spin-orbit interaction and Dressenlhaus interaction
can be regarded as particular Yang-Mills fields.
The current tensor is related to the SU(2)$\times$U(1) gauge
potential.
We have also provided a precise definition of orbital angular momentum
current.
Using the Lagrangian formalism, we have constructed
a conserved total current, which consists of a conventional bare spin current and
a non-vanishing term contributed from the Yang-Mills field.
The latter provides a microscopic
interpretation of the presence of a spin precession resulted
in the non-conservation of the bare spin current.
We have derived a general formula describing the forces acting on the spin and spin current.
We have proposed that the spin density and spin current
can be regarded as a source generating Yang-Mills fields in a similar
way as the Maxwell field generated by charge density and current.

The work is supported by NSFC grant No.10225419 and by RGC in Hong Kong.


\begin{references}

\bibitem{Wolf} S. A. Wolf, D. D. Awschalom, R. A. Buhrman, J. M. Daughton,
S. von Molnar, M. L. Roukes, A. Y. Chtchelkanova, D. M. Tresger,
Science {\bf 294}, 1488 (2001).

\bibitem{Zutic}
I. Zutic, J Fabian, and S. Das Sarma,
Rev. Mod. Phys. {\bf 76}, 323 (2004) and reference therein.

\bibitem{Hirsch} J. E. Hirsch, Phys. Rev. Lett. {\bf 83}, 1834 (1999).
M. I. Dyakonov and V. I. Perel, JETP Lett. {\bf 13}, 467 (1971).

\bibitem{Zhang}
S. Murakami, N. Nagaosa, and S. C. Zhang, Science {\bf 301} 1348-1351 (2003);
{\it ibid.}, Phys. Rev. B {\bf 69}, 235206 (2004).

\bibitem{Niu0403}
J. Sinova, D. Culcer, Q. Niu, N. A. Sinitsyn, T. Jungwirth, and A. H. MacDonald,
Phys. Rev. Lett. {\bf 92}, 126603 (2004).

\bibitem{Shen} S. Q. Shen, M. Ma, X. C. Xie, and F. C. Zhang,
Phys. Rev. Lett. {\bf 92}, 256603 (2004).

\bibitem{Inoue} J. I. Inoue, G. E. W. Bauer, and L. W. Molenkamp, Phys. Rev. {\bf B67}, 033104 (2003).

\bibitem{Halperin} E. G. Mishchenko, A. V. Shytov, and B. I. Halperin,
Phys. Rev. Lett. {\bf 93}, 226602 (2004).

\bibitem{Kato}
Y. Kato, R. C. Myers, A. C. Gossard and D. D. Awschalom, Nature {\bf 427} 50-53 (2004);
{\it ibid.}, Phys. Rev. Lett. {\bf 93}, 176601 (2004);
{\it ibid.}, Science, {\bf 306}, 1910 (2004).

\bibitem{Awschalom}
V. Shi, R. C. Myers, Y. K. Kato, W. H. Lau, A. C. Gossard and D. D. Awschalom, Nature Physics {\bf 1}, 31-35
(2005).

\bibitem{Wunderlich}
J. Wunderlich, B. K\"astner, J. Sinova, and T. Jungwirth,
Phys. Rev. Lett. {\bf 94}, 047204 (2005).

\bibitem{Niu0407}
D. Culcer, J. Sinova, N. A. Sinitsyn, T. Jungwirth, A. H. MacDonald, and Q. Niu,
Phys. Rev. Lett. {\bf 93}, 046602 (2004).

\bibitem{Rashba}
E. I. Rashba, Phys. Rev. B {\bf 68}, 241315 (2003);
E. I. Rashba, Sov. Phys. Solid State 2, 1109 (1960)].

\bibitem{Hu}
J. P. Hu, B. A. Bernevig, and C. J. Wu,
Int. J. Mod. Phys. B {\bf 17}, 5991-6000 (2003).

\bibitem{Schliemann}
J. Schliemann and D. Loss, Phys. Rev. B {\bf 69}, 165315 (2004).

\bibitem{Shen2003}
S. Q. Shen, Phys. Rev. B {\bf 70}, 081311(R) (2004);
N. A. Sinitsyn, E. M. Hankiewich, W. Teizer, and J. Sinova,
Phys. Rev. B {\bf 70}, 081312(R) (2004).

\bibitem{ZhangYang}
S. Zhang and Z. Yang, Phys. Rev. Lett. {\bf 84}, 06602 (2005).

\bibitem{Yang}
C. N. Yang and R. L. Mills, Phys. Rev. {\bf 96}, 191-195 (1954).

\bibitem{Anandan}
J. Anandan, Phys. Lett. A {\bf 138}, 347 (1989);
{\it ibid.}, Phys. Rev. Lett. {\bf 85}, 1354 (2000).

\bibitem{Darwin}
C. G. Darwin, Proc. Roy. Soc. A {\bf 118}, 634 (1928).

\bibitem{Davydov}
A. S. Davydov \emph{Quantum Mechanics}, second edition (Pergamon Press, New York, 1976).
G. Dresselhaus, Phys. Rev. {\bf 100} 580 (1955).

\bibitem{Nitta9702}
J. Nitta, T. Akazaki, H. Takayanagi, and T. Enoki,
Phys. Rev. Lett. {\bf 78}, 1335-1338 (1997).

\bibitem{Stevens}
M. J. Stevens and A. L. Smirl, R. D. R. Bhat, A. Najmaje, J. E. Sipe, and H. M. van Driel,
Phys. Rev. Lett. {\bf 90}, 136603 (2003).

\bibitem{Bernevig}
B. A. Bernevig and S. C. Zhang, Phys. Rev. Lett {\bf 96}, 106802 (2006).

\bibitem{Ohe}
J. Ohe, M. Yamamoto, T. Ohtsuki, and J. Nitta,
Phys. Rev. B {\bf 72}, 041308(R) (2005).

\bibitem{Shen05}
S. Q. Shen, Phys. Rev. Lett. {\bf 95}, 187203 (2005) .

\end{references}
\end{document}